\documentclass[]{piparticle-final}
\usepackage{graphicx}
\usepackage{amsmath}
\usepackage{upgreek}
\usepackage{epstopdf}
\newcommand{\um}{\upmu\textrm{m}}
\newcommand{\etal}{\textit{et al.{}}}
\newcommand{\sample}[1]{\mathrm{\scriptscriptstyle{#1}}}
\newcommand{\insertfigure}[4][0.49]{\begin{figure}\begin{center}%
            \includegraphics[width=#1\textwidth]{#2}
            \end{center}\caption{#4}\label{#3}\end{figure}}

\usepackage{cite}

\begin{document}
\volume{3}               
\articlenumber{030002}   
\journalyear{2011}       
\editor{L. Vi\~na}   
\reviewers{R. Gordon, Department of Electrical and Computer Engineering, \\ \mbox{}\hspace{36.5mm} University of Victoria, British Columbia, Canada.}  
\received{8 March 2011}     
\accepted{30 May 2011}   
\runningauthor{P. M. Jais \itshape{et al.}}  
\doi{030002}         

\title{Plasmon-enhanced second harmonic generation in semiconductor quantum dots close to metal nanoparticles}

\author{Pablo M. Jais,\cite{conicet,lec}\thanks{E-mail: jaisp@df.uba.ar}\hspace{0.5em}
        Catalina von Bilderling,\cite{cma,conicet}\thanks{E-mail: catalina@df.uba.ar}\hspace{0.5em}
        Andrea V. Bragas\cite{conicet,lec}\thanks{E-mail: bragas@df.uba.ar}}

\pipabstract{
We report the enhancement of the optical second harmonic signal in non-centrosymmetric semiconductor CdS quantum dots, when they are placed in close contact with isolated silver nanoparticles. The intensity enhancement is about 1000. We also show that the enhancement increases when the incoming laser frequency $\omega$ is tuned toward the spectral position of the silver plasmon at $2\omega$, proving that the silver nanoparticle modifies the nonlinear emission.
}

\maketitle

\blfootnote{
\begin{theaffiliation}{99}
   \institution{conicet} Instituto de F\'{\i}sica de Buenos Aires, CONICET, Argentina.
   \institution{lec} Laboratorio de Electr\'onica Cu\'antica, Departamento de F\'{\i}sica, Facultad de Ciencias Exactas y Naturales, Universidad de Buenos Aires, Intendente G\"uiraldes 2160, Pabell\'on I - Ciudad Universitaria, Buenos Aires, C1428EHA, Argentina.
   \institution{cma} Centro de Microscop\'{\i}as Avanzadas, Facultad de Ciencias Exactas y Naturales, Universidad de Buenos Aires, Argentina.
\end{theaffiliation}
}

\section{Introduction}
Second Harmonic Generation (SHG) is the second order nonlinear process for which two photons with the same frequency $\omega$ interact simultaneously with matter to generate a photon of frequency $2\omega$. The nonlinear response of a material to an applied field becomes evident in the presence of intense fields, as the one given by a focused pulsed laser.

Due to symmetry requirements for second order nonlinear processes, the leading dipolar term of the SHG is forbidden for centrosymmetric materials. However, this condition is relaxed at surfaces, where the symmetry is broken. On the other hand, for a small and perfectly spherical particle of any type of material under homogeneous illumination, the SHG from the surface is also zero since the geometrical shape recovers the symmetry \cite{Brudny2000}.

The presence of inhomogeneities in the electromagnetic field recovers the SHG even for symmetric materials, both from bulk and surface, as pointed out by Brudny \etal\ \cite{Brudny2000}. Those inhomogeneities may be, for instance, the consequence of a strong focusing of the field, the proximity of a surface or the presence of another particle. Besides, electronic resonances and enhancement of the field around the particle will increase the SHG and, in many cases, would make it detectable.

It is very well-known, as one of the main results of nano-optics, that the field around a metal nanoparticle is confined and enhanced when the incoming wavelength is resonant with the surface plasmons sustained by the nanoparticle. This effect is exploited in many different applications of metal nanoparticles: volume reducers for diffusion measurements \cite{Estrada2008}, optical probes for high resolution imaging \cite{Scarpettini2009,Kalkbrenner2001,Anger2006}, biosensors \cite{Haes2002}, nanoheaters \cite{Skirtach2006} and surface enhanced Raman scattering (SERS) \cite{Etchegoin2003,Etchegoin2009}, among others.

In the present paper, we show the enhancement of the SHG when non-centrosymmetric CdS quantum dots are placed in close contact with silver nanoparticles. The plasmon of the nanoparticles is resonant at the second harmonic (SH) frequency and enhances the emission. As far as we know, there are few works on the enhancement of SHG (either in semiconductor QDs or in molecules) produced by the resonant excitation of metal nanoparticles \cite{Ishifuji2006, Clark2000, Baldelli2000, Barsegova2002}.

\section{Materials and methods}
The experiment was performed in transmission, by performing spectrally-resolved photon counting (Fig.~\ref{experiment}). The laser was a KMLabs tunable modelocked Ti:Sapphire with $50\,$fs pulse width, $400\,$mW average power, $80\,$MHz repetition rate and tunable in the range 770--$805\,$nm. The resolution of the monochromator was $2\,$nm in all experiments. Silver nanoparticles (NPs), of average diameter $20\,$nm, were synthesized as in Marchi \etal\ \cite{Marchi1999}, while the CdS quantum dots (QDs), $3\,$nm in diameter, as in Frattini \etal\ \cite{Frattini2005}.

\insertfigure[0.39]{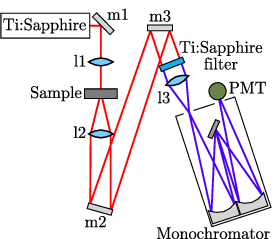}{experiment}{Experimental setup. PMT: photomultiplier tube; m1 to m3: mirrors, l1 to l3: lenses.}

The absorption spectra for both NPs and QDs in solution are shown in Fig.~\ref{samples}. The arrows mark the first and second excitonic transitions in the QDs, at $(362\pm4)\,$nm and $(425\pm15)\,$nm, obtained by the second derivative method \cite{Ekimov1993}. Figure~\ref{samples} also shows the two-photon photoluminescence (TPPL) spectrum of the QDs in solution, excited at $\lambda=780\,$nm and measured with the setup shown in Fig.~\ref{experiment}. A Stokes shift was observed in the TPPL, in accordance with the same effect reported for the (one photon) photoluminescence of CdS QDs \cite{Yu2003}.

\insertfigure{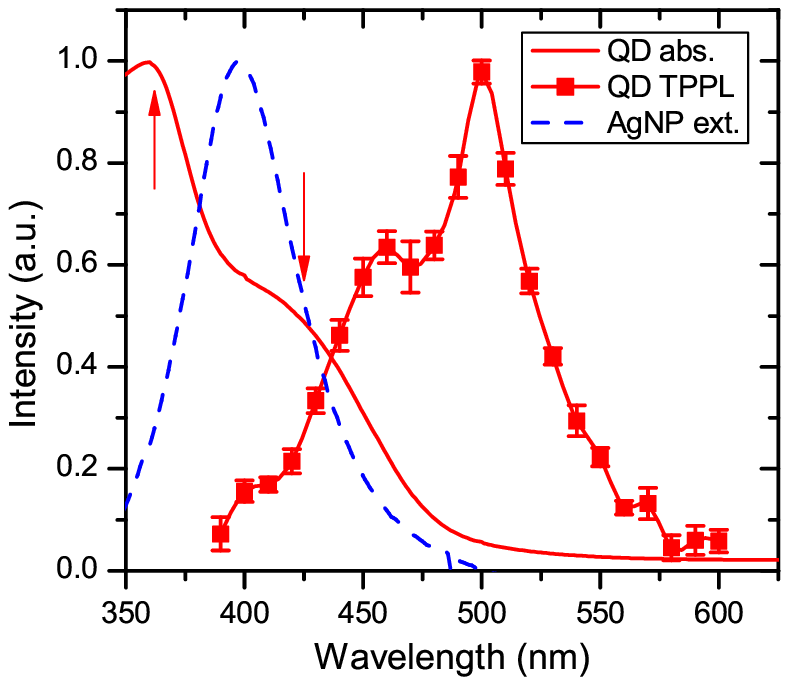}{samples}{Absorption (red line) and two-photon photoluminescence, TPPL (red squares), of the CdS QDs in ethanol solution. The dotted blue line is the extinction of the Ag NP in aqueous solution.}

NP samples were prepared by placing a drop of the NP solution on a glass substrate until the solvent evaporated. The average NP concentration was $\approx 10$ particles per $\um^2$. However, optical and SEM images revealed that the metal NPs are inhomogeneously distributed on the substrate, with a high concentration of particles in the fringe (drop boundary).

A CdS QD sample was prepared according to the same protocol. However, the QD solution has a very high concentration of aminosilanes. They form a thick layer when dried, so the QDs are immersed in an aminosilane matrix. The estimated height of this layer was $\approx 5\,\um$, and the average QD concentration was $\approx 10^5$ particles per $\um^2$ (inferred from AFM images taken at a 1:1000 diluted concentration).

For the mixed NP-QD sample, we first dried a drop of the metal NP solution, and then we deposited a drop of CdS QD solution on top of it. The resulting structure is schematically shown in Fig.~\ref{samplecartoon}.

\insertfigure[0.35]{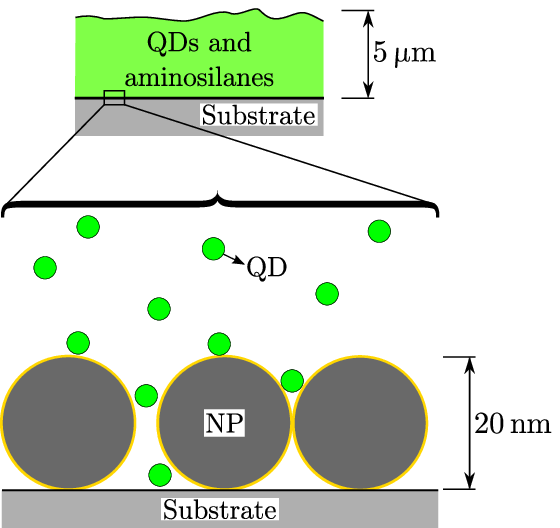}{samplecartoon}{Diagram of the NP-QD sample.}

\section{Results and discussion}
For the silver NP samples we did not detect SH signal above the noise level of the experiment (of about 200 counts per second). This result agrees with the fact that silver is a centrosymmetric material and the NPs are almost spherical, even considering that their interaction should produce a weak SH signal. On the other hand, Fig.~\ref{tppl}a shows that the CdS QD sample presents a small SH peak (as expected for a non-centrosymmetric material) and a strong TPPL signal. The central wavelength of the excitation laser for this measurement was $784\,$nm, and its FWHM was $26\,$nm.

\insertfigure{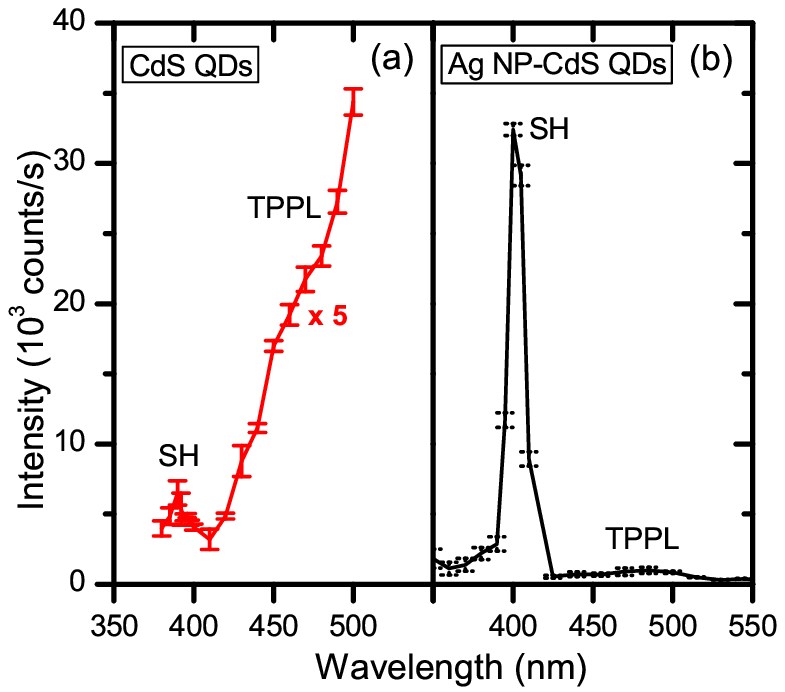}{tppl}{SH and TPPL spectra for (a) the CdS QD sample (multiplied by 5) and (b) the mixed Ag NP-CdS QD sample. Note that the SH peak was 20 times higher in the mixed sample, and the ratio between the SH and the TPPL changes drastically.}

However, a strong SH signal was measured in the mixed NP-QD sample, as shown in Fig.~\ref{tppl}b. The SHG was 20 times larger in this sample, and the SH${}/{}$TPPL ratio increased significantly. In addition, in the mixed sample, the TPPL decayed after a few seconds of laser irradiation while the SHG remained stable over time. In this case, the central wavelength was $799\,$nm, and the FWHM was also $26\,$nm.

The observations can be attributed to the enhancement of the nonlinear field around the QDs due to silver NP plasmon resonance at $2\omega$.The incoming light at $\omega$ is not enhanced by the silver NP, but creates a weak nonlinear signal at $2\omega$ in the QDs. This outgoing signal is enhanced by the field enhancement factor ($\eta$), so that the SHG intensity is enhanced by $\eta^2$. Note that in the analogous case of SERS, the incoming and outgoing signals are enhanced giving an intensity enhancement of $\eta^4$ \cite{libroetchegoin}.

The magnitude of this enhancement is not easy to calculate since the mixed NP-QD sample contains several hot-spots (usually in the concentrated NP fringe), and their intensities were wildly different. Nevertheless, the highest intensities recorded for the NP-QD sample were $\approx 10^4\,$cts/s, while the highest ones for the QD sample were $\approx 1700\,$cts/s. To estimate the enhancement, we used the analytical enhancement factor (AEF), defined as \cite{LeRu2007}
\begin{equation}
\mathrm{AEF} = \frac{I_{\sample{NP-QD}}/\rho_{\sample{NP-QD}}}{I_{\sample{QD}}/\rho_{\sample{QD}}},
\end{equation}
where $I$ and $\rho$ are the SHG intensity and the surface concentration of QDs in the corresponding samples.

We must take into consideration that there is a thick layer of QDs that are not in the enhancement region of the NPs, i.e., the measured signal in the NP-QD sample is the sum of the intensity coming from QDs close to the NPs that are enhanced, and the intensity from QDs far from the NPs that are not enhanced. If we keep only the enhanced fraction,
\begin{equation}
\mathrm{AEF} \approx \frac{I_{\sample{NP-QD}}-I_{\sample{QD}}}{I_{\sample{QD}}}\frac{\rho_{\sample{QD}}}{\tilde{\rho}_{\sample{NP-QD}}},
\end{equation}
where $\tilde{\rho}_{\sample{NP-QD}}$ is the concentration of QDs near the NPs. Since the NP enhancement decays one order of magnitude in $5\,$nm \cite{maier}, we can take $\tilde{\rho}_{\sample{NP-QD}} \approx \rho_{\sample{QD}} \frac{5\,\mathrm{nm}}{5\,\um}$. This gives AEF${}\approx 5\,10^3$. This is just an estimation of the order of magnitude of the enhancement, since the uncertainties in the thicknesses and maximum intensities are very high, but this estimation is consistent with the enhancements usually found in SERS experiments \cite{libroetchegoin}.

To study the spectral dependence of the enhancement, and gain a deeper insight into the behavior of the system, the incoming laser light was tuned while keeping the laser always on the same spot. This measurement is shown in Fig.~\ref{shg}, where the intensity is the photon flux at half the laser wavelength, normalized to the signal provided by a $40\,$fs pulse with $275\,$mW average power, according to the following formula:
\begin{equation}
I_{\mathrm{norm}} = I \frac{\delta t}{40\,\mathrm{fs}} \left(\frac{275\,\mathrm{mW}}{P}\right)^2.
\end{equation}
The signal at the SH wavelength was, for each point of Fig.~\ref{shg}, a local maximum in the spectrum.

Figure~\ref{shg} shows that the SH from the QD sample increases toward the excitonic resonance in about a factor of 2 for the whole tuning range, a value similar to the one reported by Baranov \etal\ \cite{Baranov1996}. It is worth noting that the contribution of the TPPL to this increase is much smaller than the SH signal throughout the measured range (see Fig.~\ref{tppl}). Showing a very different spectral response, the SH for the mixed NP-QD system sharply increases in more than a factor of 7 close to the resonance of the silver plasmon.

\insertfigure{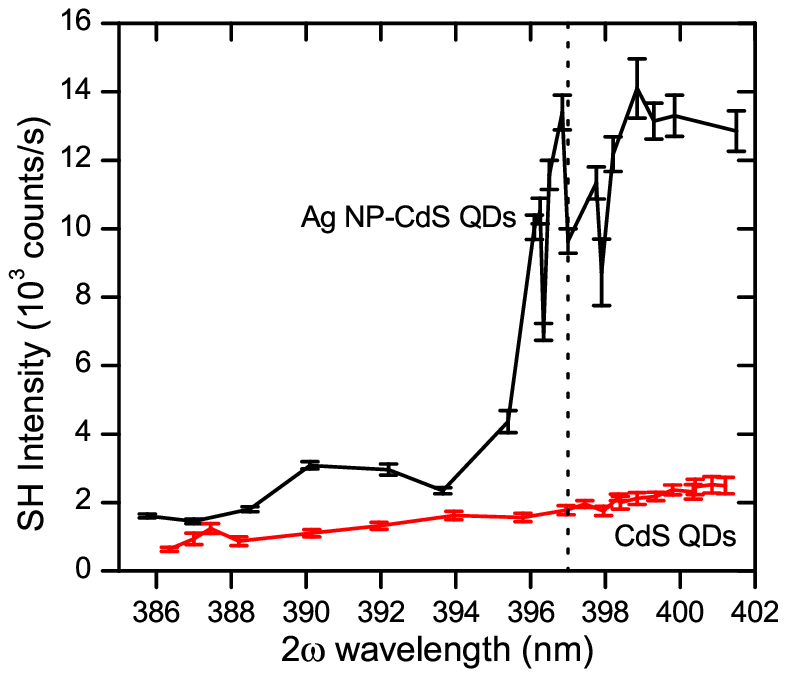}{shg}{SHG as a function of the laser excitation wavelength. The horizontal scale is the second harmonic of the incident photon. Red symbols show the SHG for the QD sample, while the black symbols show the SHG for the mixed NP-QD sample. The dotted line marks the spectral position of the maximum of the plasmon resonance.}

This measurement reinforces the hypothesis that the silver NP is modifying the nonlinear response of the QDs thanks to the resonant excitation of the silver NP plasmons. However, it must be noted that the enhancement does not exactly follow the plasmon resonance spectrum shown in Fig.~\ref{samples}. The reason for this is still unknown, but it is possible that the plasmon resonance was shifted due to the interaction among NPs. Unfortunately, the tuning range of the laser was much smaller than the spectral width of the plasmon resonance.

\section{Conclusions}
We have observed a strong enhancement ($\approx 10^3$) of the SHG when CdS QDs are mixed with silver NPs, compared with a CdS QD sample. The spectral dependence of the enhancement shows that the NP plasmons are resonantly excited by the SH emission of the QDs. These measurements are evidence that the SH enhancement is mediated by nanoparticle plasmons. This effect can be applied to significantly improve traditional applications of SH measurements such as the study of surface deposition and orientation of molecules, among others.

\begin{acknowledgements}
We thank Claudia Marchi for her help with the NP synthesis and Nora Pellegri for providing the QDs. We also thank Alejandro Fainstein for the insightful discussions. This work was supported by the University of Buenos Aires under Grant X010 and by ANPCYT under Grant PICT 14209.
\end{acknowledgements}

\end{document}